\documentclass[preprint,12pt]{aastex}

\newcommand{\as}{\mbox{\arcsec}}

\def\lsim {$\rlap{\raise.4ex\hbox{$<$}}\lower.55ex\hbox{$\sim$}\,$}

\def\c17o{$\rm C^{17}O$}
\def\dc18o{$\rm C^{18}O$}

\begin{document}

\title {\bf Molecular Line Profiles from a Core Forming in a Turbulent Cloud }
\author {Jeong-Eun Lee}
\affil{\it Department of Astronomy and Space Science, Astrophysical
Research Center for the Structure and Evolution of the Cosmos,
Sejong University, 98 Gunja-Dong, Gwangjin-Gu, Seoul 143-747, Korea}
\email{jelee@sejong.ac.kr}
\and
\author {Jongsoo Kim}
\affil{\it Korea Astronomy and Space Science Institute, Hwaam-Dong,
Yuseong-Gu, Daejeon 305-348, South Korea}
\email{jskim@kasi.re.kr}

\begin{abstract}
We calculate the evolution of molecular line profiles of HCO$^+$ and C$^{18}$O toward a dense core that
is forming inside a magnetized turbulent molecular cloud.  Features of the profiles can be affected more significantly
by coupled velocity and abundance structures in the outer region than those in the
inner dense part of the core.
The velocity structure at large radii is dominated by a turbulent flow nearby
and accretion shocks onto the core, which resulting in the variation between
inward and outward motions during the evolution of the core.  The chemical abundance structure is
significantly affected by the depletion of molecules in the central region with high density and low temperature.  During the evolution of the core, the asymmetry of line profiles easily changes from blue to red, and vice versa.
According to our study, the observed reversed (red) asymmetry toward some starless
cores could be interpreted as an intrinsic result of
outward motion in the outer region of a dense core, which is embedded in a turbulent
environment and still grows in density at the center.
\end{abstract}


\section{Introduction}

Stars form inside very dense molecular clouds by gravitational collapse.
Starless dense cores, which have been considered as potential sites of future
star formation, have been used to study the initial conditions of star formation.
There is, however, no straightforward method to determine whether specific cores
will actually form stars.
Observations tell us the mass or density profile of a core, which provides clues to whether
the core is gravitationally bound or unstable to collapse to form a star
in the future.

Walker et al. (1986) and Zhou et al. (1993) interpreted the asymmetry of line
profiles with blue stronger peaks ({\it blue asymmetry}), which have been seen in
optically thick self-absorbed lines, as an infall signature.
Gregersen \& Evans (2000) and Lee et al. (2001) have surveyed dense
starless cores with molecular lines of HCO$^+$ 3$-$2 and CS 2$-$1.
The infall signature is seen in about 35\% and 50\% of the observed starless
cores of each survey, respectively.
The infall signature, however, occurs in much larger scales
across the face of a core
opposed to the collapsing center expected from the inside-out collapse
model (Shu 1977), indicative of an overall inward motion.

This overall inward motion in starless dense cores could be considered as a result of
a supercritical (or slightly subcritical) ambipolar diffusion model in Ciolek \& Basu (2000).
The inward velocity profile in L1544, the canonical starless dense core, is able to be reproduced
reasonably well with the model (Caselli et al.~2002).
However, in addition to the blue asymmetry, the {\it reversed asymmetry}, which has
a red stronger peak in the self-absorption feature, is also seen toward many starless cores.
This reversed asymmetry cannot be caused by outflows, which can be developed only
after an accretion disk forms around a young stellar object.
However, by definition, a starless core can not harbor the young stellar object.
Other studies claim that the reversed asymmetry might be caused by
rotation (Redman et al. 2004; Lee et al. 2007) or oscillation
(Keto et al. 2006; Maret et al. 2007; Aguti et al. 2007) of dense cores.

As described above, the interpretation of the velocity profiles toward starless dense cores is still controversial since the formation process of dense cores is not  well understood.
Here, we suggest a possible explanation of the reversed red asymmetry observed toward some starless cores with the theory based on turbulence since it involves not only inward but also outward motions
in a core formed from the density fluctuations of a larger turbulent cloud (see, e.g., Mac Low \& Kessen 2004).
In addition to the dynamical process, the chemical process in a
dense core can also affect molecular line profiles (Rawlings \& Yates 2001,
Lee et al. 2003, 2004, 2005). Therefore, observed line profiles must be
interpreted based on both the dynamics and chemistry of a core.

In this Letter, we combine dynamics and chemistry in order to study
the evolution of molecular line profiles toward a dense core forming within
a large turbulent cloud. We show that
the velocity structure at the outer part of the core can significantly affect
molecular line profiles.  Indeed, simulated line profiles exhibit
a mixture of infall and outflow signatures as a function of time.
Thus a snap shot of core conditions of a single object cannot
provide a full accounting of core formation and evolution toward collapse.
This Letter is organized as follow.  The dynamical and chemical models
of a core are presented in \S 2, results, and summary and discussion are given in
\S 3 and \S 4, respectively.

\section{Dynamical and Chemical models of a Core}

\subsection{Dynamical Evolution of a Core}

What we are interested in this study is to follow up the evolution of molecular line profiles
of a core forming in a turbulent molecular cloud.
For this purpose, we perform a couple of three-dimensional numerical experiments of a self-gravitating, isothermal, turbulent, magnetized molecular cloud. 
In fact, we integrate, as a function of time, the isothermal MHD and Poisson equations.
The number of cells used in the experiments is $512^3$.  A typical core is covered with more than $32^3$ cells.

There are two parameters in the numerical experiments:
one is the plasma $\beta$, which is the ratio of gas pressure to
magnetic pressure, and the other is the Jeans number $J$, which is the
ratio of the one dimensional size of a computational box $L$, to the
Jeans length $L_J$.  A mass-to-flux ratio normalized with its
critical value $\mu$, can be written in terms of these two parameters
$\mu=\pi J\sqrt{\beta/2}$, where we use the critical mass-to-flux ratio
$(M/\phi)_{\rm cr} = (4\pi^2G)^{-1/2}$ in Nagano \& Nagamura (1978).
Initially a uniform molecular cloud is threaded with a uniform
magnetic field.  In order to generate a turbulent flow and keep the level
of the turbulence inside the molecular cloud, we add velocity perturbations
generated in the Fourier space with wavelengths that span from half to full size of the
box.  We adjust the input rate of the kinetic energy so that the root-mean-square
sonic Mach number $M_s$ of a turbulent flow becomes 10 (Mac Low 1999; Stone et al. 1998).
Once a fully saturated turbulent flow is generated, we turn on the self-gravity of gas and follow up the evolution of a densest core formed in the cloud.
If $\beta=0.1$ and $J=4$, then the $\mu$ value is 2.8.  So the initial molecular cloud is in a
highly supersonic ($M_s>>1$), Jeans unstable ($J>1$), and magnetically super-critical
($\mu>1$) state.

Since the combined isothermal MHD and Poisson equations can be written
in dimensionless form with the two parameters, we can select any units of, for
example, length, time, and mass in such a way that they are scalable with
each other.  However there is one more additional parameter $M_s$, 
which is used to 
fix these units.  If Larson's relation between the sizes and velocity
dispersions of clouds (Larson 1981) holds for our model cloud, then the Mach
number, which is the velocity dispersion of our model cloud, can
determine its corresponding size.  In fact, a Larson-type relation 
of the form $M_s=5(L/1 {\rm pc})^{1/2}$, if we take 0.5 as the power index 
(for example, Myers 1983), is used.
Then density and magnetic field strength could be
determined from relations, $n = 500(J/(L/1 {\rm pc}))^2$ cm$^{-3}$,
and $B=0.205(n/\beta)^{1/2}$ $\mu$G (see also equation~(7) and equation~(8)
in V\'azquez-Semadeni et al.~2005).  So if we choose
$\beta=0.1$, $J=4$, and $M_s=10$, the size of our computational box, the
initial number density, and the initial magnetic field strength become 4 pc,
500 cm$^{-3}$, and 14.5 $\mu$G, respectively.

Our dynamical model has a few limitations such as turning on
self-gravity after generating a fully saturated turbulent flow,
driving turbulent flows by adding velocity fields generated in the Fourier
space, not taking into account ambipolar diffusion,
and assuming the isothermal condition. However, they are common
in numerical models of a turbulent molecular cloud
(see, for example, Klessen et al.~2000; Ostriker et al.~2001;
Li et al.~2004; Vazquez-Semadeni et al.~2005). Due to the
limited space, we only mention the effects of the latter two
limitations on our results. If ambipolar diffusion is included
in our model, then the ambipolar diffusion process is more
important in the central part of a core where the degree of
ionization is lower than that of the outer part of the core. Since our
synthesized line profiles are more sensitive to the conditions
at the outer layer (see section 3), our main
results are not likely to be significantly affected by
ambipolar diffusion.
We also take the isothermal
approximation which enables us not to solve the energy equation
with cooling and heating processes. To take into account
the cooling and heating processes of gas realistically entails a heavy
calculation because molecular cooling lines are usually
optically thick. 
When line profiles are, however, synthesized, we calculate two cases with a constant temperature and temperature variations of gas and dust inside the core (see section 3). We present the latter case only because the line features from both cases are not much different from each other.

\subsection{Chemical Evolution and the HCO$^+$ Line Evolution}

In order to calculate the chemical evolution in the core picked-up from
the dynamical simulation as described in the previous subsection,
we adopt the evolutionary chemical model developed by Lee et al. (2004),
in particular, the model before collapse begins.
In the prestellar phase where
this paper focuses on, the model calculates the chemical evolution at each grid point
as density grows with time.
The chemical network in the model has been updated to include more recent results
on the binding energy of N$_2$ (\"Oberg et al. 2005) and the photodesorption yield of CO
(\"Oberg et al. 2007).
For the initial molecular abundances of the coupled evolution between dynamics
and chemistry, we calculate the chemical evolution in a constant density of
$5\times10^3$ cm$^{-3}$ with the same initial atomic abundances as used in
Lee et al.~(2004) for $t=5\times10^5$ years.
For the chemical calculation, the dense core is assumed to be surrounded with
material of $A_V =1.0$ mag, where CO is self-shielded.
For consistency, adopting the methods used in Lee et al. (2004), we also calculate
the dust and gas temperatures with the interstellar radiation field attenuated
by $A_V =1.0$ mag.
The time step for data dump of the dynamical simulation is $4\times10^4$ years, during which
the chemistry is calculated based on constant density and temperatures given at a specific dump
time, and the result of the chemical evolution is
used as initial conditions for the next dump time.

We use the Monte-Carlo method (Choi et al. 1995) to perform the radiative
transfer calculation for the HCO$^+$ and C$^{18}$O lines.
The collision rates for HCO$^+$ and CO are adopted from Flower (1999) and
Flower \& Launay (1985).
For the simulation of the line profiles, we assume the distance to the
model core of 250 pc and the beam sizes for the Caltech Submillimeter Observatory
10.4 m telescope.
We also include a constant microturbulent velocity dispersion of 0.2 km s$^{-1}$.
We model the HCO$^+$ 3$-$2 and 4$-$3 transitions,
which have been used to study infall motion in low mass cores with and
without young stellar objects
(Gregersen et al. 1997; Gregersen \& Evans 2000).
The optically thin C$^{18}$O 3$-$2 line is also modeled to be compared to
the HCO$^+$ line profiles.

\section{Results}

We use a one-dimensional (radial) chemical network coupled with a radiative
transfer model to simulate line profiles.  For this purpose we have to reduce the
three-dimensional density and velocity fields into one-dimensional radial
distributions using the following procedures.  First, we find the position of
a density peak of the whole computational domain, which is taken as the origin of the radial
coordinate.  Second, we identify a core as a connected region whose density
is larger than $n_{\rm H_2}=10^3$ cm$^{-3}$.  Third, we subtract a center-of-mass
velocity of the core from its velocity field. The three-dimensional center-of-velocity of the core ranges from 0.2 km sec$^{-1}$ to 2.0 km sec$^{-1}$
Fourth, we take the average of the density and radial velocity at each spherical shell with the width of a computational cell. At this point we have to mention the effect of the radial
averaging that we have taken in the above. The averaging
certainly smoothed out density and velocity variations at an
outer layer affected by, for example, a shock. So the line
profiles observed towards a specific direction
may show much stronger
variation than the line profiles that are obtained after taking
the radial averaging.

In Figure 1, we plot the density and velocity
profiles as a function of the radial distance from the center of the core.
The two digit numbers are the elapsed time in units of 40,000 years after
turning on self-gravity.  We can clearly see from density profiles that the
central density of the core is increasing as time goes on.
The power index of density profiles in the outer region is
increasing from shallower than -2 at an earlier time to steeper than -2 at a later time.
At a given radial distance, the density and velocity dispersions are much smaller than the radially averaged density and velocity, respectively.

The radial velocity profiles are quite interesting.  In the central part of the core, most of
the profiles have negative but very small velocity, which means that the central part of core is slowly
collapsing.
In the outer part of the core, the profiles have both positive
and negative signs and their amplitudes are larger than the thermal sound
speed.  The large-amplitude velocity profiles are due to not only the large-scale turbulent flow nearby the core but also accretion shocks onto the
core.  Even though we do not show radial density and velocity profiles
from a simulation with a weak initial field strength, $\beta=1.0$,
the large-amplitude velocity profiles in the outer part of a core are also found.  However, infall velocities in the
central part of the core have a bit larger (but still have a subsonic speed)
than ones of the case with $\beta=0.1$.

Figure 2 shows the comparison between dust and gas temperatures calculated based on
the dust continuum radiative transfer and the balance between heating and cooling of gas,
respectively. The gas temperature is decoupled from the dust temperature at densities
$\le 5\times 10^4$ cm$^{-3}$, where gas-grain collisions is not dominant,
and has peaks at the surface of the model core due to the photoelectric heating.

The evolution of the HCO$^+$ radial abundance calculated based
on the density and temperature evolution is presented in Figure 3.
As expected, at small radii, HCO$^+$ is significantly depleted from the gas phase
as density grows because CO becomes frozen-out onto grain surfaces (Lee et al. 2003, 2004 and
references therein) in such dense and cold conditions.
The abundance drops at the core surface due to the
dissociative recombination of HCO$^+$ with electrons.
As described in \S 2.2, the chemical evolution is calculated at each grid point without
following each gas parcel (that is, assuming the quasi-static evolution)
due to the following reasons.
In the inner region of the core, the static evolution is a good
approximation since the velocity is much smaller than the sound speed. In the outer part of the core, however, densities are low, and thus
the chemical time scale is too long to be affected by the movement
of gas parcels. For example, an infalling gas parcel with the infall
velocity of 0.5 km s$^{-1}$ at the surface moves inward only $\sim$0.02 pc
for 40,000 years,
and the densities of two positions are not very different.
This assumption is also well supported in Figure 3.
The abundance of HCO$^+$ varies
significantly at small radii, but at radii greater than 0.1 pc, where
velocities are possibly greater than the sound speed, the abundance
does not vary much with time.

Figure 4 presents the evolution of the HCO$^+$ 3$-$2, 4$-$3, and
C$^{18}$O 3$-$2 line profiles toward the core center.
The two digit number in each panel represents the elapse time in units of 40,000 years
after turning on the self-gravity.  The first upper left panel shows
line profiles at $1\times 10^6$
($25\times 40,000$) years and the time of the second panel is increased by $8\times 10^4$
($2\times 40,000$) years from the first one, and so on.  
Although the density grows with time, the C$^{18}$O 3$-$2 line
becomes weaker since CO is depleted from gas due to its
freeze-out onto grain surfaces.
The most important trend seen in this study is that the
asymmetry of the HCO$^+$ 3$-$2 line profile
mostly depends on the velocity structure at the outer part of the core.
In the comparison of line profiles with the velocity structures
(Figure 1b), if the velocity at large radii shows infall motion(e.g., at the time of
37),
the HCO$^+$ line has the infall asymmetry with a blue stronger peak.
On the other hand, the line profile represents the reversed asymmetry with
a red stronger peak when the velocity at large radii shows outward motion
(e.g., at the times of 25, 31, and 49).
This result is caused by the combination of the abundance and
velocity profiles; HCO$^+$ is depleted at small radii so that
the HCO$^+$ 3$-$2 lines effectively form at rather large radii in the core,
where velocities are dominated by the turbulent flow and accretion shocks.
The kinetic temperature also increases outward. As a result, the excitation temperature
of HCO$^+$ 3$-$2 peaks around 0.1 pc.
Even the broad line feature seen at time 43, which may be
interpreted as an infall at the core center in actual observations, is
caused by the motion at large radii of the core rather than that
occurring at small radii.

In the comparison of the HCO$^+$ 4$-$3 line (gray) with the 3$-$2 line (black
solid) in Figure 4, the 3$-$2 line shows the reversed asymmetry while the blue
asymmetry appears in the HCO$^+$ 4$-$3 line at time 25 and 29
(the opposite feature at time 35 and 47)
although the difference between blue and red peak strengths is not significant.
This varying asymmetry in different transitions of a molecule has been
observed (e.g., L1157 in Gregersen et al. 1997; B68 in Maret et al. 2007).
In addition, the relative strength between two peaks of the HCO$^+$ 3$-$2 line
is reduced in the 4$-$3 line at the times of 31 and 39.
These trends are featured since the two transitions trace different radii of
the core, for example, at time 25, the 4$-$3 line traces a
deeper region, which has inward motion, while the 3$-$2 line traces the less
dense outer region, which shows outward motion.

We assume that our model core of a radius 0.25 pc is at a distance 250 pc.  
Its angular size is about 200\as\ .  
Since physical conditions of the core at around 0.1 pc ($\sim$ 80\as\ ) mainly
contribute to the line formation, a telescope beam size that is not larger than 
the angular size of the core does not significantly affect the line intensity and 
shape shown in Figure~4.

\section{Summary and Discussion}

We traced, through a high-resolution numerical experiment, the evolution of
a core forming in a turbulent molecular cloud that is in a Jeans unstable and
magnetically supercritical state.  The core is slowly contracting at the central
part with a subthermal speed.  Due to the influence of accretion shocks and a
nearby
turbulent flow, velocity fields in the outer region of the core can easily exceed
the thermal speed, and show not only the contracting but also sometimes expanding
motions. Based on the evolutionary profiles of density and gas and dust temperatures,
we calculated the evolution of the HCO$^{+}$ abundance.
Once the central density of the core
is larger than $10^5~{\rm cm}^{-3}$, the HCO$^{+}$ abundance profiles peak at
outer regions because of the depletion of HCO$^{+}$ at the central
part of the core (see Figure~3).
We also calculated the evolution of HCO$^{+}$ 3$-$2,
4$-$3, and C$^{18}$O 3$-$2 line profiles that are coupled to the density, kinetic temperature,
and abundance structures.
The most interesting result in this work is that the HCO$^{+}$ line profiles are
dominantly affected by the velocity profiles at
the outer regions. This is the consequence of both the HCO$^{+}$
abundance and velocity profiles that have peaks at the outer regions.

We showed that the molecular line profiles of a core, which results from the coupled
velocity and chemical structures, vary with time.
To understand the formation process of cores, survey observations of many cores in different evolutionary states
are crucial.
Mapping observations of individual cores with a high resolution enables us to know
the complex velocity structures of the cores.
The radial variation of the velocity field as seen in B68 may be an intrinsic part of star formation,
and the blue asymmetry is not always a direct signature of collapse in
starless dense cores (Keto et al. 2006).

\acknowledgments
We are very grateful to Ted Bergin for valuable comments.
This work was supported by the Korea Science and Engineering Foundation
under a cooperative agreement with the Astrophysical Research Canter for the
Structure and Evolution of the Cosmos.

\clearpage

\begin{figure}
\figurenum{1}
\epsscale{1.0}
\plotone{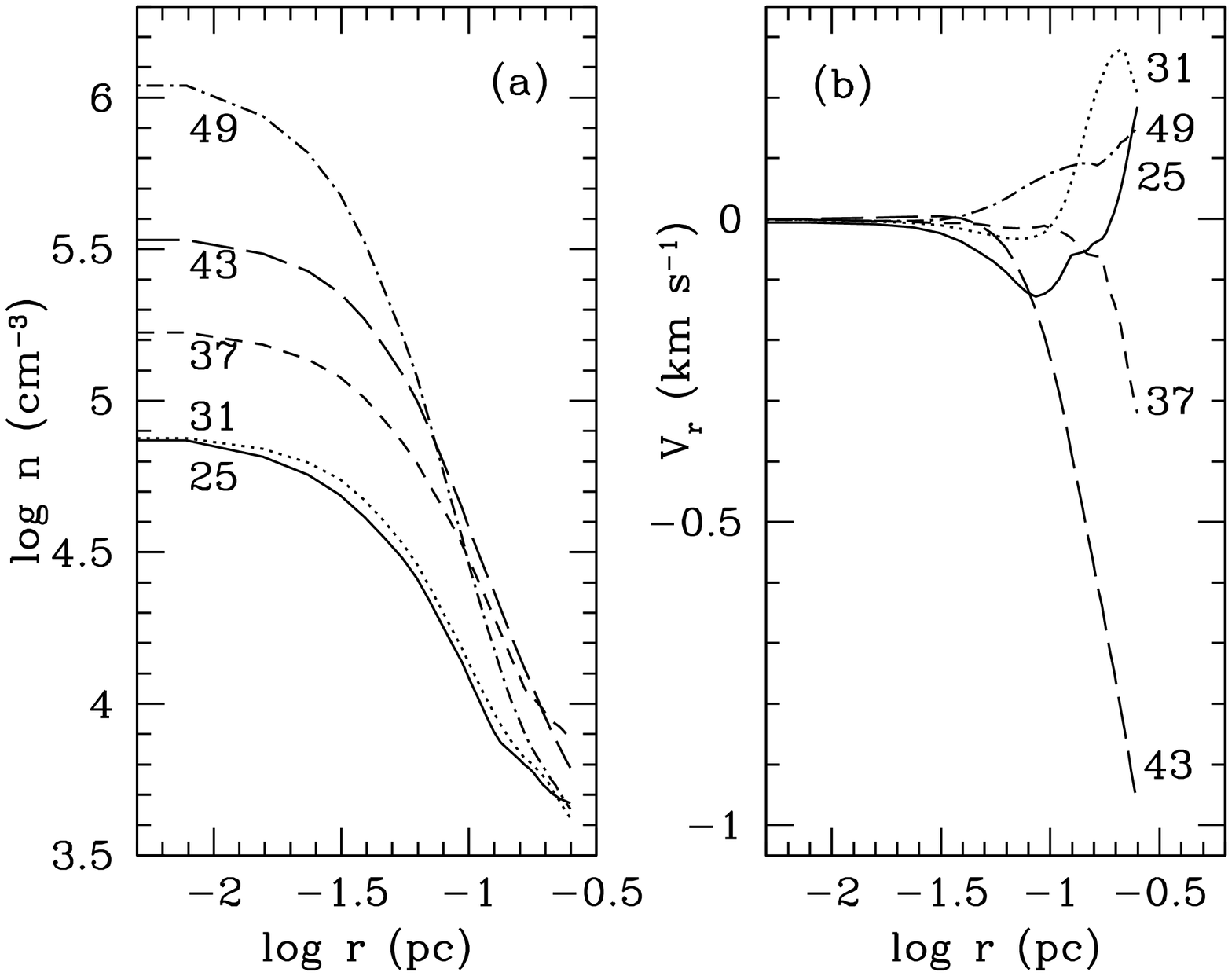}
\caption{
The evolution of density and velocity profiles.
Two digit numbers represent time in units of 40,000 years.
}
\end{figure}

\clearpage

\begin{figure}
\figurenum{2}
\epsscale{1.0}
\plotone{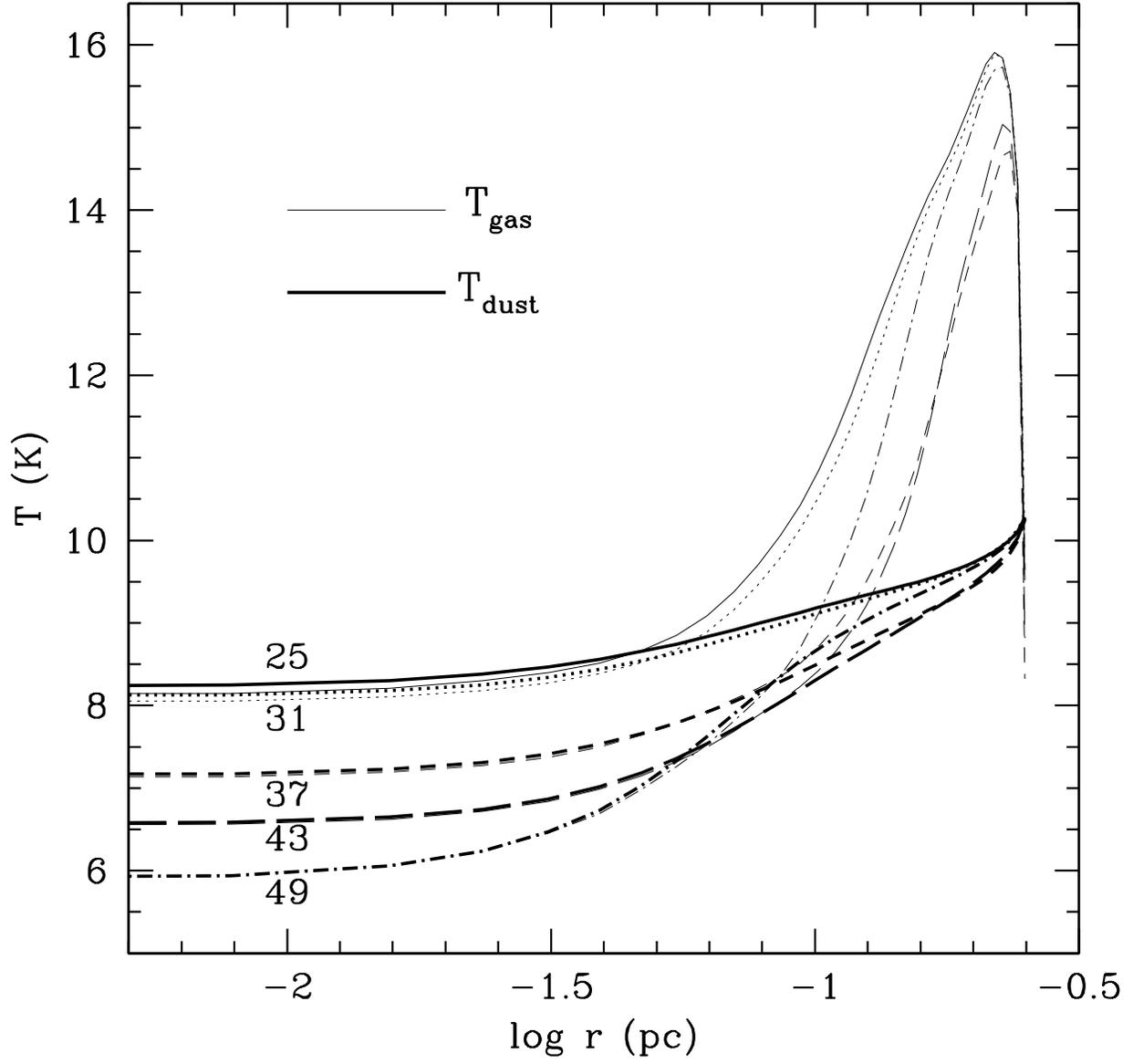}
\caption{
The evolution of the gas (thin lines) and dust (thick lines) temperature profiles.
The two digit numbers represent time in units of 40,000 years.}
\end{figure}

\clearpage

\begin{figure}
\figurenum{3}
\epsscale{1.0}
\plotone{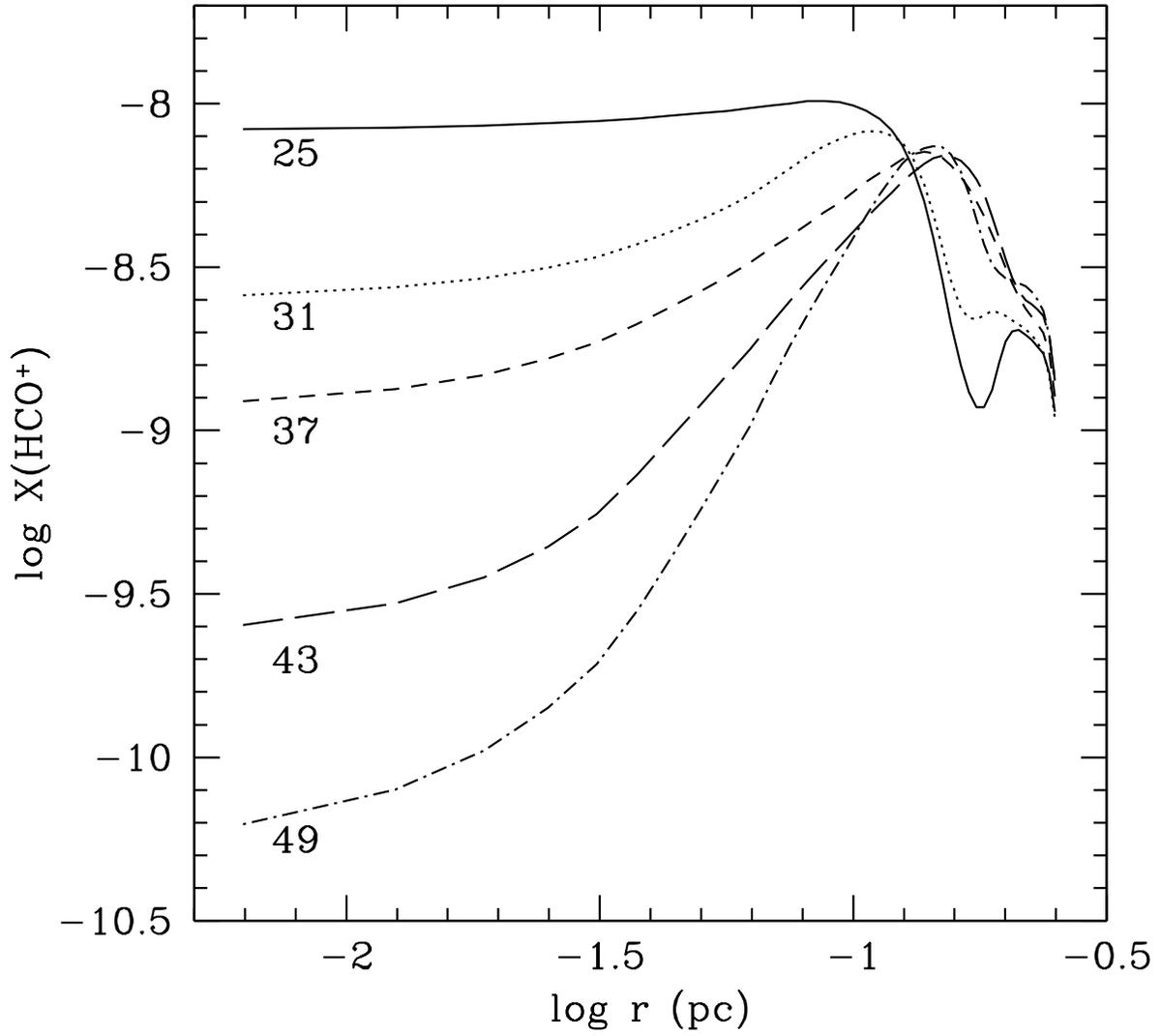}
\caption{
The evolution of the HCO$^+$ abundance profiles. 
The two digit numbers represent time in units of 40,000 years.
}
\end{figure}

\clearpage

\begin{figure}
\figurenum{4}
\epsscale{1.0}
\plotone{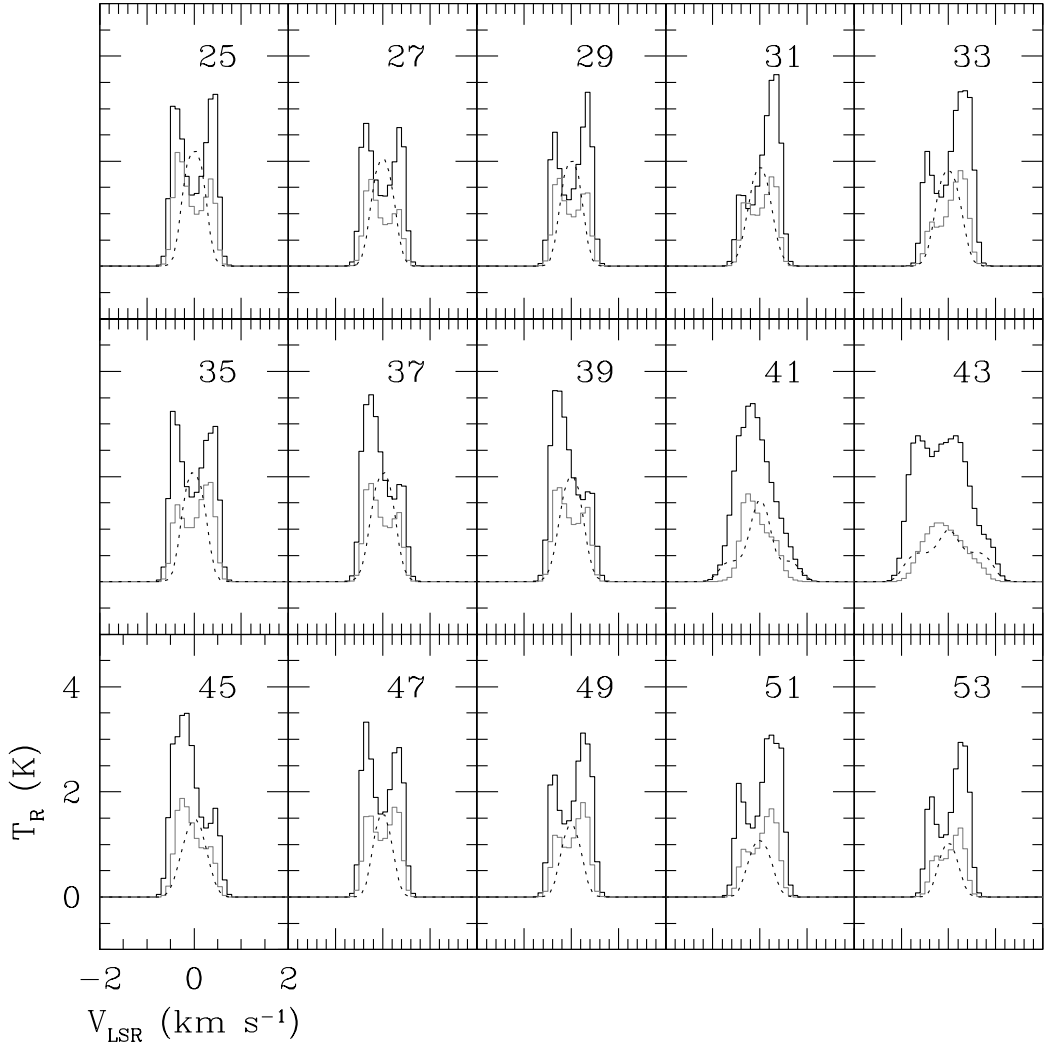}
\caption{
The evolution of molecular line profiles. The black and gray solid
histograms represent the HCO$^+$ 3$-$2 and 4$-$3 line profiles,
respectively. The dotted lines show the C$^{18}$O 3$-$2 lines.
The beam sizes are 26\as\ for HCO$^+$ 3$-$2 and
C$^{18}$O 3$-$2 and 20\as\ for HCO$^+$ 4$-$3, respectively.
The angular size of our model core with a radius of 0.25 pc at a distance of 250 pc 
is around 200\as\ . The two digit numbers represent time in units of 40,000 years.
}
\end{figure}

\clearpage

\end{document}